\documentclass[aip,reprint,floatfix,superscriptaddress,showpacs,amsfonts,amssymb,amsmath,preprintnumbers]{revtex4-1}

\usepackage{graphicx}
\usepackage{float, amsmath,verbatim,amssymb,stmaryrd}
\usepackage{epsfig}

\begin{document}

%\preprint{}

\title{Laser absorption via QED cascades in counter propagating laser pulses}

\author{T. Grismayer }
\email{thomas.grismayer@ist.utl.pt}
\address{GoLP/Instituto de Plasmas e Fus\~ao Nuclear, Instituto Superior T\'ecnico-Universidade de Lisboa, Lisbon, Portugal}
\author{M. Vranic}
\address{GoLP/Instituto de Plasmas e Fus\~ao Nuclear, Instituto Superior T\'ecnico-Universidade de Lisboa, Lisbon, Portugal}
\author{J. L. Martins}
\address{GoLP/Instituto de Plasmas e Fus\~ao Nuclear, Instituto Superior T\'ecnico-Universidade de Lisboa, Lisbon, Portugal}
\author{R. A. Fonseca}
\address{GoLP/Instituto de Plasmas e Fus\~ao Nuclear, Instituto Superior T\'ecnico-Universidade de Lisboa, Lisbon, Portugal}
\address{DCTI/ISCTE Instituto Universit\'{a}rio de Lisboa, 1649-026 Lisboa, Portugal}
\author{L. O. Silva}
\email{luis.silva@ist.utl.pt}
\address{GoLP/Instituto de Plasmas e Fus\~ao Nuclear, Instituto Superior T\'ecnico-Universidade de Lisboa, Lisbon, Portugal}

\date{\today}

\begin{abstract}

A model for laser light absorption in electron-positron plasmas self-consistently created via QED cascades is described. The laser energy is mainly absorbed due to hard photon emission via nonlinear Compton scattering. The degree of absorption depends on the laser intensity and the pulse duration. The QED cascades are studied with multi-dimensional particle-in-cell simulations complemented by a QED module and a macro-particle merging algorithm that allows to handle the exponential growth of the number of particles. Results range from moderate-intensity regimes ($\mathrm{\sim 10~PW}$) where the laser absorption is negligible, to extreme intensities ( $\mathrm{> 100~PW}$) where the degree of absorption reaches $80\%$. Our study demonstrates good agreement between the analytical model and simulations. The expected properties of the hard photon emission and the generated pair-plasma are investigated, and the experimental signatures for near-future laser facilities are discussed.

 \end{abstract}


%\keywords{}

\maketitle
%Intro_

\section{Introduction}

Relativistic electron-positron pair plasmas are tightly related to extreme astrophysical objects such as pulsar magnetospheres, blazar jets or even gamma-ray bursts. Due to the inherent difficulties of studying these remote objects it is extremely desirable to study dense pair plasmas in the laboratory, both for fundamental purposes and astrophysical applications\cite{Res_plas_astro}. The recent spectacular rise in laser intensities such as the development pursued on the HERCULES laser\cite{Yanovsky}, accompanied by the ongoing construction of new laser facilities such as ELI \cite{ELI} or the Vulcan 20 PW Project \cite{Vulcan} will place intensities above $10^{23}$ W/cm$^2$ within reach, i.e., generating electric fields of TV/m and magnetic fields of few GigaGauss\cite{RevModPhysdipiazza}. The magnitude of these electromagnetic fields overlaps with the estimated fields of magnetic white dwarfs and milliseconds pulsars\cite{Res_plas_astro}. Producing pair plasmas in ultra strong fields may demonstrate that we can mimic the conditions appropriate to the atmospheres of these extreme astrophysical environments in terrestrial laboratories. As pulsars efficiently convert the large scale Poynting flux to gamma-rays, the laboratory analog of a pulsar is expected to efficiently convert optical light into gamma-rays\cite{Gruzinov}. The pair creation in such energy density environments is not caused by the Schwinger mechanism\cite{Schwinger} but by the decay of high energy photons (gamma rays) in intense fields. This process usually leads to quantum electrodynamics (QED) cascades, as the pairs created re-emit hard photons that decay anew in pairs, eventually resulting in an electron-positron-photon plasma. Gruzinov\cite{Gruzinov} has estimated that to concert the entire laser pulse into gamma-rays, the number density of pairs should be $n \sim 10^{23}L_{\mathrm{PW}}^{1/2}\lambda_{\mu}^{-2} \mathrm{cm^{-3}}$ which reaches one to few orders of magnitude the density of solids for 10-100 PW lasers. In astrophysics, this density is known as the Goldreich-Julian density which is capable of noticeably altering and depleting the external fields. QED cascades also referenced as electronic or electromagnetic showers \cite {Landau_showers, Akhiezer, Anguelov_Vankov, Erber} when the external field is purely magnetic, have also been theoretically studied in different electromagnetic configurations \cite{Bulanov_pairsvaccuum, Bell_Kirk_2lasers, Bulanov2013, Gelfer, Jirka, Tamburini}. Notably, Bell \& Kirk \cite{model1bell} suggested a judicious configuration comprising two circularly polarized counter propagating lasers with some seed electrons in the interaction region to initiate the cascade. They predicted prolific pair production for intensities approaching $10^{24}$ W/cm$^2$ for a $\mu$m wavelength laser. Among many analytical and numerical works devoted to QED cascades in ultra-intense laser pulses, only a few of them\cite{Nerush_laserlimit, Ridgers_invited, Brady_2012, Fedotov_cascade} have drawn their attention on the laser absorption which turns to be crucial to achieve high pair plasma densities and efficient optical photons conversion to gamma rays. In fact, analytical predictions are difficult to achieve for such complicated scenarios and the extreme multiplicity of pairs and photons in the cascade development requires tremendous computational power to allow multi-dimensional simulations. 

In this article, we present a study of QED cascades in counter propagating laser pulses of 10-100 PW, leveraging on the original setup proposed by Bell \& Kirk \cite{model1bell}. We first recall the initial development of the cascades that are characterised by a single parameter, the growth rate which depend itself on the laser intensity and polarisation. The core of the article is devoted to the laser absorption for linearly polarised lasers. In order to address numerically this late time of the cascade where the colossal multiplicity has led to a photon-pair plasma whose kinetic energy becomes comparable with the initial energy of the pulse, we resort on the coupling of our QED-PIC framework with a particle merging algorithm which permits, despite the exponential growth of the particles, to keep the number of PIC particles approximatively constant in the simulation box.  The absorption efficiency as a function of the initial intensity is investigated in multi-dimensional geometries and a phenomenological model is found to support to the simulations results. Besides detailed predictions regarding the photons and pair-plasma self-consistently generated, we demonstrate the conditions required to achieve strong laser absorption regimes in upcoming laser facilities. 

\begin{figure*}%\centering
\includegraphics[width=1\textwidth]{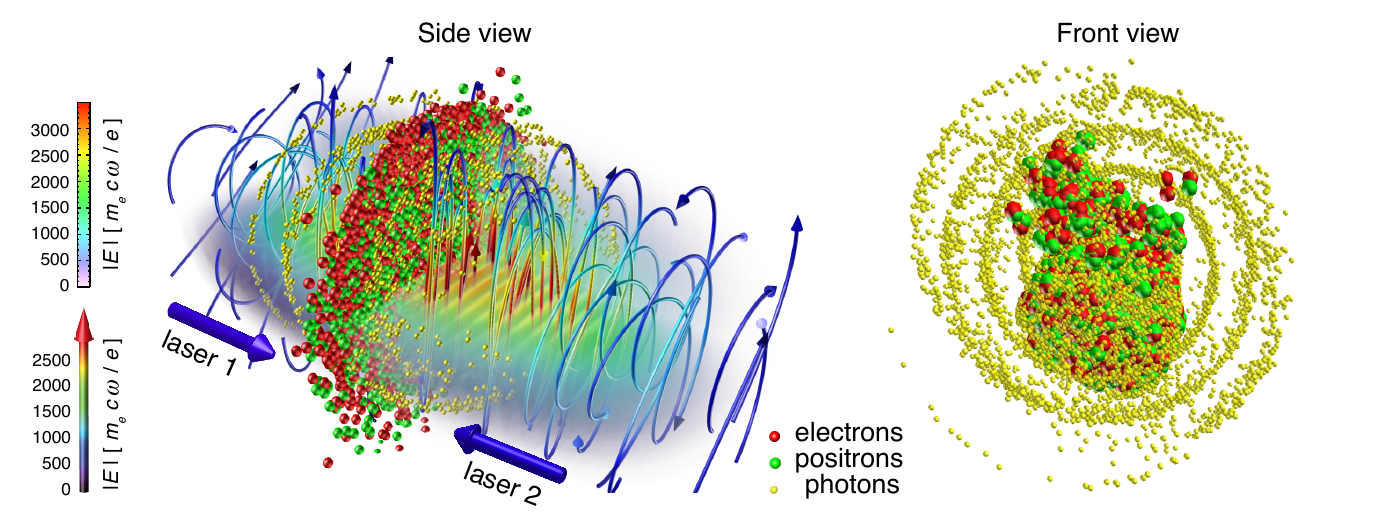}
\caption{Side view and front view of the development of a QED cascades in 3D. The magnitude of the electric field resulting from the beating of the two laser pulses is represented by the coloured bar. The curves lines with arrows represent the electric field lines. The electrons, positrons and photons are respectively displayed in red, green and yellow. The particles shown only represent a small fraction of the particles of the simulation.} 
\label{fig:cascade3d}
\end{figure*}

\section{QED-PIC simulations}

Our exploration relies on a QED module, part of our particle-in-cell (PIC) code OSIRIS 3.0 \cite{OSIRIS}, which includes real photon emission from an electron or a positron, and decay of photons into pairs, known as the Breit-Wheeler process.  The differential probability rates describing these processes can be found in \cite{pair_rate1, pair_rate2, pair_rate3, Ritus_thesis, Erber}. The key results of this article are strongly grounded in QED-PIC simulations. This novel approach has been pioneered by several groups \cite{Gremillet, lobet, Ridgers_quantumRRnew, Elkina_rot,Ridgers_solid, Nerush_laserlimit, Bell_Kirk_MC, Gonoskov2015review} and our implementation follows the standard methodology. Many QED-PIC simulations have been performed in order to benchmark our module with previous results \cite{Gremillet, Elkina_rot, Nerush_laserlimit, Basmakhov, tang}. One of the challenges of QED-PIC simulations is the emergence of a vast number of particles that makes the simulations demanding in terms of memory requirements. In the scenarios we are interested in this article, i.e., at the focus of ultra intense lasers, the very localised regions of extremely strong field can easily produce vast numbers of electron-positron pairs even starting from just one seed electron \cite{Elkina_rot, Nerush_laserlimit} or one photon \cite{Basmakhov,tang} leading to an exponential growth of the number of particles and photons. The enormous amount of particles being created severely hinders PIC simulation performance. This issue can be sorted out with the use of particle merging algorithm which allows to resample the 6D phase space with different weighted macro-particles. We have developed a merging algorithm\cite{Vranic_merging} that preserves the total energy, momentum and charge as well as the particle phase space distribution generalising previous attempts to merge particles that considered only the conservation of some of the physical quantities \cite{Timokhin, Lapenta_new, Lapenta_old}.

\section{Development of the cascade}
\label{sec:dev_cas}

We first address the development of the QED cascades. This plays an important role in the laser absorption features. As a meter of fact, the growth rate of the cascade sets a determinant time scale for the cascading process. As discussed elsewhere, this depends on the laser polarisation, and on the laser intensity. For the sake of completeness, we review here some of the key results presented in the reference\cite{grismayer}.

\subsection{Effect of the polarisation}

To motivate our discussion we first present simulations where we have explored configurations of colliding laser pulses whose polarization can either be linear or circular. In our simulations, all laser pulses have a $\lambda_0 = 1 \mu \mathrm{m}$ central wavelength, and the same spatio-temporal envelope functions, with differences in the fast-oscillating components that will be presented separately for different polarisations. The envelope function is transversally a Gaussian  with a focal spot $\sigma_0$ of 3.2 $\mu \mathrm{m}$, while the slope of the temporal profile is given by $10\tau^3-15\tau^4+6\tau^5$ ($\tau \in [0,1]$), $\tau=t/\tau_0$ where $\tau_0=32~$fs is the pulse duration at FWHM in the fields and the pulse is symmetric with respect to the point of maximum intensity.  The laser pulses are initialised 20 $\mu \mathrm{m}$ away from each other. The focal plane for both lasers is located at half distance between their envelope centers, and several test electrons are placed there to seed the cascade. The simulation box is composed of $3000\times 1200$ cells and $3000\times 1200\times 1200$ cells for 2D and 3D, respectively. After extensive convergence tests, the spatial resolution was set to $dx=dy=dz=0.1~c/\omega_0$ and $dt= 0.001~ \omega_0^{-1}$ ($\omega_0=k_0c=2\pi c/\lambda_0$). The three-dimensional development of the cascade in the case of circularly polarised laser is shown in Fig.\ref{fig:cascade3d}. By examining the geometry of the standing waves, we can develop an intuitive picture on how the particles are accelerated, and hence predict which configuration will be the optimal. For a given $a_0$, a way to define the optimal configuration consists in finding the one that offers on average the highest values of the quantum invariant \cite{Ritus_thesis, Erber}
\begin{equation}
\chi = \frac{1}{E_S}\sqrt{(\gamma\vec{E}+\vec{u}\times \vec{B})^2-(\vec{u}\cdot\vec{E})^2}
\end{equation}
where $E_S = m^2c^3/e\hbar$ and with $\vec{u}=\vec{p}/mc$. It should be emphasised that radiation reaction in intense fields modifies the orbits of particles \cite{Esirkepov} and can lead to anomalous radiative trapping \cite{Gonoskov2014} which we omit in our following analysis but which is self-consistently included in our simulations. 

Setup 1 (lp-lp) consists of two linearly polarised lasers where the phase and polarisation are defined by 
\begin{equation}
\vec{a}_\pm=(0,\pm a_0 \sin(k_0x \mp \omega_0t),0),  
\end{equation}
where the indexes ``$+$'' and ``$-$'' respectively denote a wave propagating in the positive and in the negative $x$ direction. $a_0=eE_0/m\omega_0c$ is the Lorentz-invariant parameter, related to the intensity $I$ by $a_0=0.85(I\lambda_0^2/10^{18}\textrm{Wcm}^{-2})^{1/2}$ and $E_0$ the peak electric field strength. This results in a standing wave where $E_y=2a_0 \cos (k_0x) \sin(\omega_0t),  B_z=-2a_0 \sin (k_0x) \cos (\omega_0t)$ where the fields amplitude are expressed in units of $m\omega_0c/e$. The dynamics of the particles is determined by the electric or the magnetic field depending on the phase within the temporal cycle \cite{Basmakhov, Esirkepov}. The electric field accelerates the pairs in the $y$ direction, while the magnetic field $B_z$ can rotate them and produce $p_x$ ensuring a perpendicular momentum component to both $\vec{E}$ and $\vec{B}$. The rise in $p_x$ gradually increases $\chi_e$ until a photon is radiated. The most probable locations to create pairs or hard photons are precisely $\lambda_0/4$ and $3\lambda_0/4$ \cite{Esirkepov}. For a particle born at rest, $\chi_e$ oscillates approximatively twice per laser period with a maximum on the order of $2a_0^2/a_S$ where $a_S=mc^2/\hbar\omega_0$ is the normalised Schwinger field \cite{Schwinger}. The cascade develops mostly around the bunching locations and is characterised by a growth rate that possesses an oscillating component at $2\omega_0$.  

Setup 2 (cw-cp) is formed by a clockwise and a counter-clockwise polarised laser:  
\begin{equation}
\vec{a}_\pm=(0,a_0\cos(\omega_0t\pm k_0x),\pm a_0\sin (\omega_0t\pm k_0x)),
\end{equation}
where $a_0\simeq 0.6(I\lambda_0^2/10^{18}\textrm{Wcm}^{-2})^{1/2}$. The components $E_y$ and $B_z$ are anew the same but $E_z=2a_0 \sin (k_0x) \sin(\omega_0t), B_y=-2a_0 \cos (k_0x) \cos (\omega_0t)$. This setup consists in a rotating field structure and the dynamics of the particles has been already studied \cite{model1bell, Elkina_rot, Fedotov_cascade}. The advantage lies in the direction of the fields that is constantly changing, and the particles are not required to move in $x$ to enter a region where  $\vec{E}$ and $\vec{B}$ are perpendicular to their momentum.  The particle acceleration is stronger in the regions of high electric field, so the highest electron momenta are obtained where the electric field is maximum. This then leads to higher $\chi_e$ and the cascade develops in the region of strong electric field (precisely in the node $B=0$ \cite{model1bell}) producing a plasma wheel as shown in Fig.\ref{fig:cascade3d}. At this particular position, the parameter $\chi_e$ can reach a maximal value of $2a_0^2/a_S$ \cite{Elkina_rot}.

From the description of the two configurations, the setup 1 can produce the highest values of $\chi_e$ ($\chi_e > 2a_0^2/a_S$) but only for particles born in a specific phase of the standing wave. The majority of the particles are sloshing back and forth between the electric and magnetic zone which results in lower average $\chi_e$ in comparison with setup 3. The efficiency of the cascade setups can be more accurately assessed by calculating its growth rate $\Gamma$. 

\subsection{Theoretical models}

\subsubsection{Circular polarization}
 The case of a uniform rotating electric field constitutes a good approximation of the standing wave field produced in the setup 2\cite{model1bell}. The advantage of this setup is that the cascade develops mostly in one spot $x=\pi/2$, which allows us to assume a time-dependent field. It has been shown\cite{grismayer} that the equation governing the time evolution of the number of pairs growing in the cascade is
\begin{equation}
\label{eq:intdiff}
\frac{dn_{p}}{dt}=2\int_0^tdt'\int d\chi_{\gamma}n_{p}(t')\frac{d^2P}{dt'd\chi_{\gamma}}W_{p}e^{-W_{p}(t-t')},
\end{equation}
where the pairs follow a fluid-like behaviour which can be described through an average energy $\bar{\gamma}$ and an average quantum parameter $\bar{\chi_e}$. The differential probability rate for photon emission $d^2P/dt'd\chi_{\gamma}$ depends thus on $\bar{\gamma}$, $\bar{\chi}_e$ and $\chi_{\gamma}$. The photon decay rate (or the pair emission probability rate) can be considered as constant which permits us to write $W_p=W_p(\chi_{\gamma},\epsilon_{\gamma})$ with $\epsilon_{\gamma}=\bar{\gamma}\chi_{\gamma}/\bar{\chi}_e$. Eq.(\ref{eq:intdiff}) can be solved using the Laplace transform and calculating growth rate correspond to solve the zeros 
\begin{equation}
\label{eq:charac}
s-2\int_0^{\bar{\chi_e}} d\chi_{\gamma}\frac{\frac{d^2P}{dt'd\chi_{\gamma}}W_{p}}{s+W_p} = 0
\end{equation}
 In the limit $\bar{\chi}_e\ll 1$, the pair creation probability can be approximated by \cite{Erber,Ritus_thesis} $W_p\simeq(3\pi/50)(\alpha/\tau_c)e^{-8/3\chi_{\gamma}}\chi_{\gamma}/\epsilon_{\gamma}$ and $d^2P/dtd\chi_{\gamma}\simeq \sqrt{2/3\pi}(\alpha/\tau_c)e^{-\delta}/(\delta^{1/2}\bar{\chi}_e\bar{\gamma}$) with $\delta=2\chi_{\gamma}/(3\bar{\chi}_e(\bar{\chi}_e-\chi_{\gamma}))$, $\tau_c=\hbar/mc^2$ and $\alpha= e^2/\hbar c$. We start from an assumption (which is verified by the result) that in the limit $\bar{\chi}_e\ll 1$, $W_p(\chi_{\gamma})\ll s$, hence the zeros of $s$ corresponding to a growing exponential ($\Gamma=s^{+}$) are given by 
\begin{eqnarray}
\label{eq:lowa0}
\nonumber
\Gamma_{\chi_e \ll 1 } &\simeq& \left(2\int_0^{\bar{\chi_e}} d\chi_{\gamma} \frac{d^2P}{dtd\chi_{\gamma}}W_p\right)^{1/2} \\
&\simeq& \frac{1}{5}\left(\frac{2\pi}{3}\right)^{1/4}\frac{\alpha}{\tau_c}\frac{\bar{\chi}_e e^{-8/3\bar{\chi}_e}}{\bar{\gamma}}.
\end{eqnarray}
The last step consists in finding how $\bar{\gamma}$ and $\bar{\chi}_e$ depend on $a_0=eE_0/m\omega_0c$. In a rotating field mocking the beating of two $1 \mu m$ lasers \cite{Bell_Kirk_2lasers, Elkina_rot, model1bell}, $\vec{a}=a_r[\cos(\omega_0t),\sin(\omega_0t)]$ ($a_r= 2a_0$), it is clear that $\Gamma \ll \omega_0$ for $\bar{\chi}_e\ll 1$. Thus $\bar{\gamma}$ and $\bar{\chi_e}$ can be approximated by their average values over a laser cycle: for $a_r \gg 1$ (neglecting the quantum recoil) one finds $\bar{\gamma}\simeq \langle \gamma \rangle = 4a_r/\pi$ and $\bar{\chi}_e\simeq \langle \chi_e \rangle = a_r^2/a_S$. Interestingly we see that cascades can develop below the threshold suggested by Fedotov \cite {Fedotov_cascade}($a_r > \alpha a_S$). In the other limit, $\bar{\chi_e}\gg 1$, $W_{\gamma}$ and $W_p$ have a similar asymptotic expression and the result of Fedotov \cite {Fedotov_cascade}, 
\begin{equation}
\Gamma_{\chi_e \gg 1 } \sim W_p \sim W_{\gamma}(\bar{\chi}_e,\bar{\gamma})
\end{equation} 
is consistent with Eq.(\ref{eq:charac}). In this limit, where the recoil cannot be omitted, $\Gamma \gg \omega_0$ and the values of $\bar{\gamma}$ and $\bar{\chi}_e$ can be evaluated as \cite{Fedotov_cascade} $\bar{\gamma}\sim \gamma(t=W_{\gamma}^{-1})\simeq\mu^{3/4}\sqrt{a_S}$ and $\bar{\chi}_e\sim \chi_e(t=W_{\gamma}^{-1})\simeq 1.24\mu^{3/2}$ with $\mu= a_r/(\alpha a_S)$. 

\subsubsection{Linear polarisation}
If we replace the factor $e^{-8/3\bar{\chi}_e}$ by $K_{1/3}^2(4/3\bar{\chi}_e)8/(3\pi\bar{\chi}_e)$ to account for a good estimate of the probability of pair production \cite{Erber} for all $\chi$ and lowering the value of $\chi_e$ as $\bar{\chi}_{e,l} = \bar{\chi}_e/4 = a_0^2/a_S$, one can find an empirical fit for the growth rate in the case of linear polarisation
\begin{equation}
\Gamma_l \sim \frac{8}{15\pi}\left(\frac{2\pi}{3}\right)^{1/4}\frac{\alpha}{\tau_c\bar{\gamma}}K_{1/3}^2\left(\frac{4}{3\bar{\chi}_{e,l}}\right),
\label{growthlinear}
\end{equation}
where $K_{1/3}$ denotes the modified Bessel function of the second kind and $\bar{\gamma}$ is kept to be same as for the circular polarisation.  

\subsubsection{Comparisons with simulations results}
The analytical results have been compared with PIC-QED simulations\cite{grismayer}. The growth rate given by Eq.(\ref{eq:lowa0}) and the numerical solution of Eq.(\ref{eq:charac}) are in good agreement with the rotating field simulation results in the limit of their validity. As expected, the growth rates in the cw-cc setup match those of the rotating field configuration. This growth rate is the highest of the two configurations for a fixed $a_0$. The seeding with electrons of the setup cw-cc turns out to be difficult. The reason is that an efficient growth happens only in the regions around the maximum of the electric field. By starting a cascade with only a few electrons, it is not guaranteed that they will enter such a region. The simulations in reference\cite{grismayer} also confirms the $2\omega_0$ oscillating component of the lp-lp growth rate. There is no appreciable difference in the growth rate of the 2D and 3D simulations for linearly polarised lasers. The empirical expression for the growth rate in the case of linearly polarised lasers has also been shown to be a valuable estimate. 

\begin{figure*}[t!]
\includegraphics[width=1\textwidth]{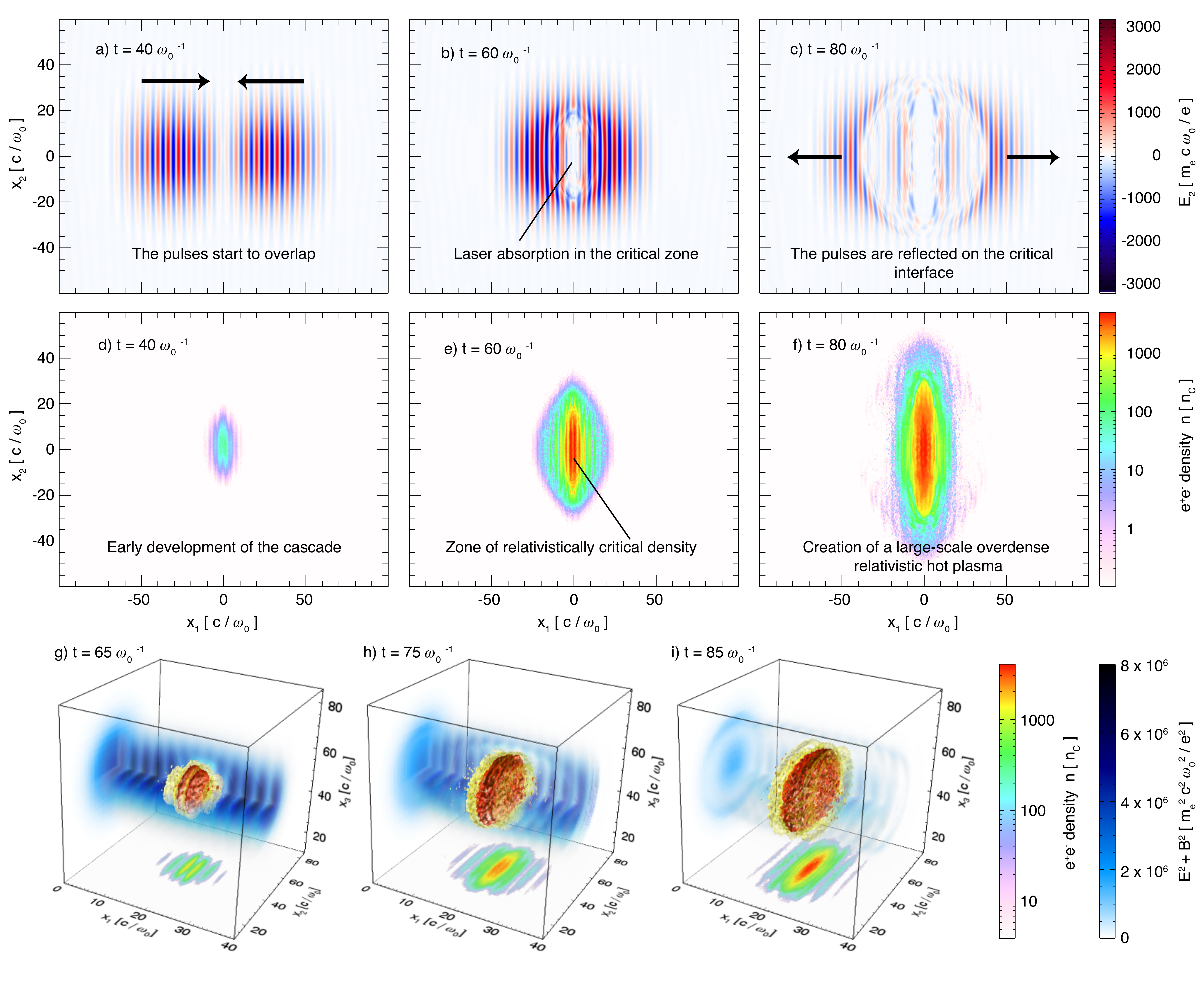}
\caption{Illustration of laser absorption in 2D and 3D cascades with two linearly polarised lasers with $a_0=2000$. a) - c) Laser electric field in 2D simulation for 3 different times; d)-f) Pair plasma density at the same instants of time as above; g) - h) Electromagnetic energy density, and pair plasma density from a 3D simulation. Iso-surfaces correspond to $n=750~n_c$ (dark orange) and $n=250~n_c$ (light yellow). Note that the lasers are further apart in the beginning of the 3D simulations, such that $t=0$ in 2D corresponds to $t=20~\omega_0^{-1}$ in 3D.} 
\label{absorption_2d3d}
\end{figure*}

\begin{figure*}
\includegraphics[width=1\textwidth]{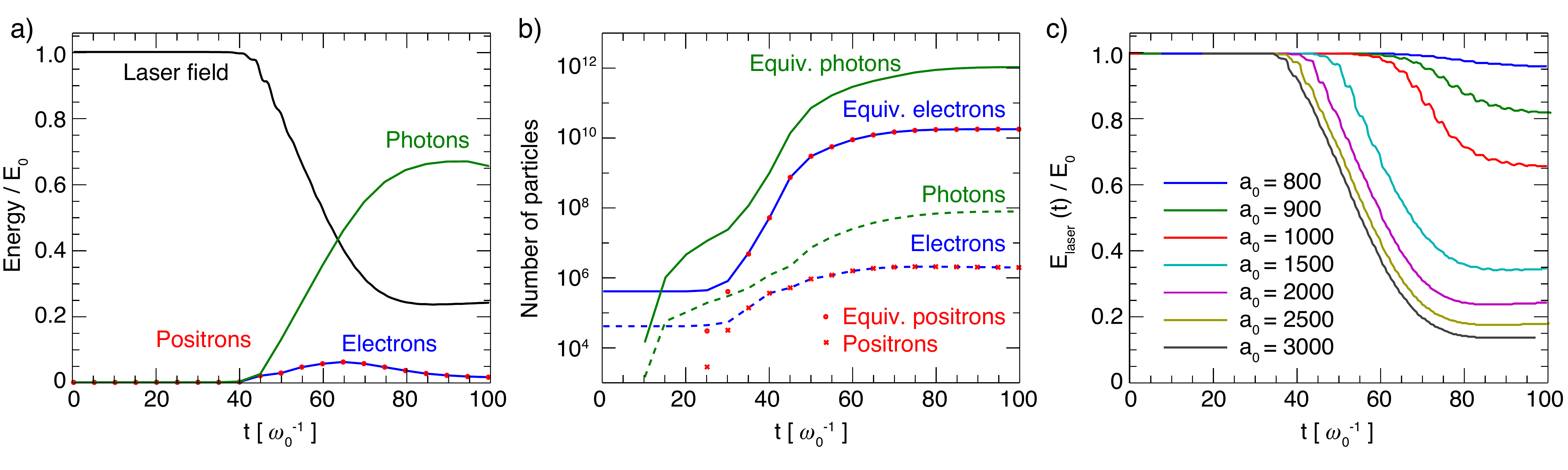}
\caption{a) Energy of the electromagnetic field, photons and pairs as a function of time for $a_0 = 2000$. $E_0$ denotes the initial energy of the laser pulses; b) Number of particles (electron, positron and photons) as a function of time for $a_0 = 2000$. The dashed curves show the real number of PIC particles and the slide lines the equivalent number of particles without the merging algorithm; c) Energy of the electromagnetic field as a function of time for different initial intensities.}
\label{absortion_time}
\end{figure*}

\section{Laser absorption}

\subsection{Classical and quantum absorption}

In non-linear electrodynamics, probabilities are calculated by taking into account a non-perturbative coupling between the pairs (or energetic photons) and the background intense field. For the processes of non linear Compton scattering or pair production to occur, it requires the simultaneous absorption of many laser photons within the formation region. These absorbed photons correspond to the least amount of laser four-momentum needed to bring the four-momenta of the particle with mass on shell \cite{meuren_thesis, meuren_article}. Without dwelling on complicated details of these QED processes, one can approximatively write that the total number of photons absorbed \cite{meuren_thesis, meuren_article} is: $n = n_{\textrm{quantum}}+ n_{\textrm{classical}}$ where $n_{\textrm{quantum}} \sim a_0/\chi$ and $n_{\textrm{classical}} \sim a_0^3/\chi$. For example, in the case of pair production, the hard photon decays into a pair at a certain laser phase $\phi^{*}$. The formation phase is very small $\delta \phi \sim \chi/a_0$ such that the decay happens almost instantaneously inside a constant crossed field \cite{Ritus_thesis}. After the creation of the pair at $\phi^{*}$, the electron and the positron propagate classically until they leave eventually the laser field. The number of photons $n_{\textrm{quantum}}$ corresponds therefore to the number of photons absorbed in the formation region where the pair is created whereas $n_{\textrm{classical}}$ represents the number of photons that the pair can asymptotically absorb. Since quantum absorption is neglected in the numerical implementation of the QED processes, it is crucial to first estimate the amount of energy related to this absorption mechanism. For every hard photon or pair created, the laser pulses are depleted approximatively by $n_{\textrm{quantum}}$ photons, i.e., an energy of $n_{\textrm{quantum}}\hbar \omega_0$. This quantity has to be multiplied by the number of times $M$ that the QED processes occurred during the cascades. Let us consider that $M \sim (10-100)N_0e^{\Gamma \tau}$, where $N_0$ is the number of initial electrons to seed the cascade, $\tau$ is the minimum between the characteristic pulse duration ($\tau_0$) and the time for the laser to reach strong absorption $t_a$. The factor (10-100) accounts for the fact that the number of hard photons exceeds the number of pairs created. The ratio $\rho$ between the energy corresponding to quantum absorption and the initial laser energy ($\mathcal{E}_L \simeq cE_0^2\sigma_0^2\tau_0/4\pi$) is 
\begin{equation}
\rho \sim \frac{M}{\chi a_0 a_Sn_c\sigma_0^2c\tau_0},
\end{equation}
where $n_c$ represents the critical density associated to the laser frequency $\omega_0$. If we consider an optical laser, linearly polarised, with the parameters discussed in the last section and a cascade seeded with a critical density target with a volume of 1 $\mathrm{\mu m^3}$, then for $a_0 \sim 1000$, $\Gamma / \omega_0 \sim 0.25$ and $M \sim 10^{16}-10^{17}$. If we set up $\chi$ to be on the order of unity, then $\rho \sim 0.003$ which remains almost negligible. As it will be shown in the next section, the maximum intensity we considered for the laser absorption study is $I =  1.2\times10^{25}$ W/cm$^2$, corresponding to $a_0 = 3000$. In this case, $M \sim 10^{18}$ and we obtain $\rho \sim 0.08$. Within this range of intensities, the assumption of a negligible quantum laser absorption seems thus reasonable.

\subsection{Simulations for linearly polarised pulses}
Self-consistent laser absorption in QED cascades has been first numerically studied by Nerush\cite{Nerush_laserlimit}. In an identical setup consisting of two counter propagating linearly polarised laser pulses ($I =  1.2\times10^{24}$ W/cm$^2$, 100 fs at $1/e^2$, $\lambda = 0.8 \mathrm{\mu m}$), the authors observed an absorption of about 40$\%$ of the initial laser energy into gamma-rays and electron-positron pairs. We have conducted a series of 2D and 3D simulations to study the efficiency of laser absorption when the intensity of the pulses (or $a_0$) is varied. We restrict this study for linearly polarised lasers since this is likely to be the polarisation that will be available for future facilities such as ELI \cite{ELI} or the Vulcan 20 PW Project \cite{Vulcan} and because it has been shown that this polarisation favours the seeding go the cascade\cite{grismayer}. The cascade is seeded by a $0.35\times0.35~\mu \mathrm{m}^2$ ($0.65\times0.35\times4~\mu \mathrm{m}^3$ for 3D) target located equidistant to the two pulses. The initial density of the target is chosen to be $n_0 = 10 n_c$ equivalent to liquid hydrogen. In 2D, the simulation box was $300\times120~ c^2/\omega_0^2$, resolved with $30000\times12000$ cells ($\sim$ 630 points per laser wavelength) and $dt=5\times10^{-3}~\omega_0^{-1}$ ($\sim$ 1260 points per laser period). In 3D, the simulation box is  $40\times96\times96~ c^3/\omega_0^3$, with $400\times960\times960$ cells and $dt=10^{-2}~\omega_0^{-1}$.

Before discussing how the absorption depends on $a_0$, we shall illustrate the three constitutive stages of the seeded QED cascades that are depicted in Fig. \ref{absorption_2d3d}:
\begin{enumerate}
\item The two pulses start to overlap and the produced standing wave allows for the development of the cascade, characterised by an exponential growth of the number of pairs and photons. At this stage, shown in the insets a), d) and g) of Fig. \ref{absorption_2d3d}, the energy contained in the pair and photon population is negligible compared to the initial laser energy. 
\item The produced pair plasma reaches the relativistic critical density and one observes a depletion of the laser in the critical zone, see insets b), e) and h) of Fig. \ref{absorption_2d3d}.
\item If the time to reach the relativistic critical density arises before a time comparable with the pulse duration (which is the case for $a_0 = 2000$) a fraction of the laser pulse is reflected on the critical interface. The reflection of the pulses is displayed by the black arrows in Fig. \ref{absorption_2d3d}-c.
\end{enumerate}
\begin{figure*}[t!]
\includegraphics[width=\textwidth]{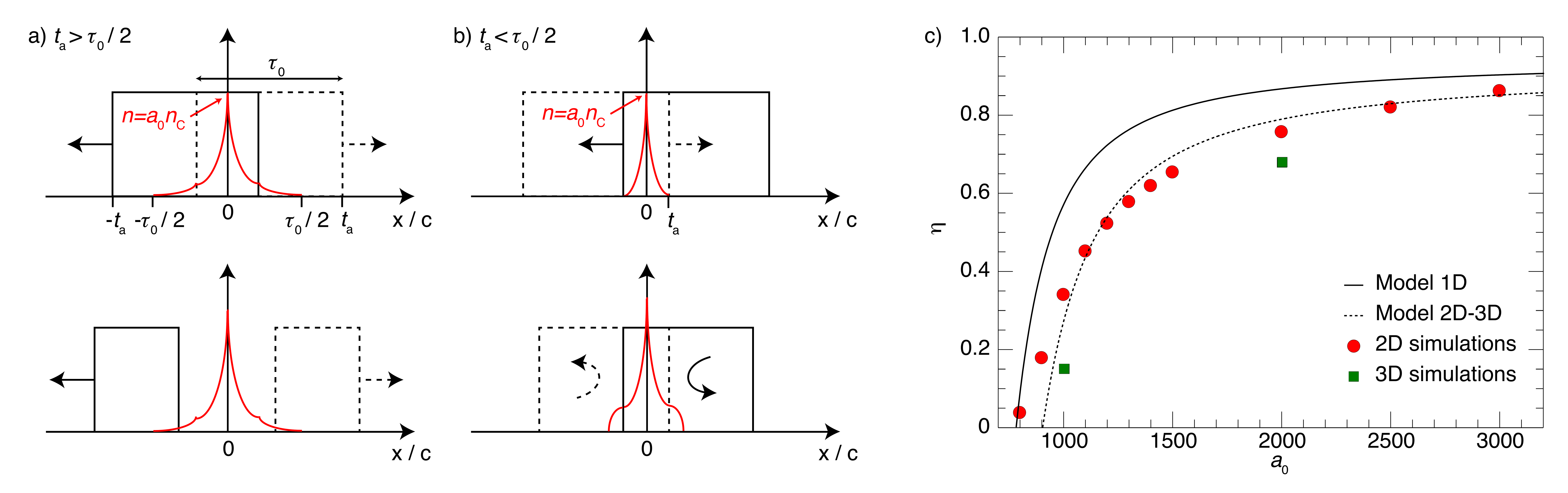}
\caption{a) upper inset: cartoon representing two flat top laser pulses overlapping at $t = t_a > \tau_0/2$ (the pulse going to the right in dashed line and the one going to the left in solid line); at this instant, the pair plasma density is depicted by the red curve and reaches the relativistic critical density at $x=0$. a) lower inset: the part of the pulses contained in the critical overlapping zone has been fully depleted and the head part of the pulses which did not interact with the plasma are moving outwards from the critical zone. b) upper inset: cartoon representing  squared laser pulses overlapping at $t = t_a < \tau_0/2$; at this instant, the pair plasma density is depicted by the red curve and reaches the relativistic critical density at $x=0$. b) lower inset: the tail of the two pulses cannot enter the critical zone and is being reflected at the critical surface forming a new standing wave. c) Laser absorption $\eta$ as a function of $a_0$; the red circles and the green square represent respectively the 2D and 3D simulations results; Eq.(\ref{eq:absorption}) is shown in solid line and in dashed line for the 1D and the 2D model respectively.}
\label{cartoon}
\end{figure*} 
The time evolution of the different components (laser, pairs and photons) of the cascades is shown in Fig. \ref{absortion_time} a-b. The inset b) of Fig. \ref{absortion_time} depicts the growth of the real number of PIC particles (pairs and photons) in the simulation as well as the equivalent number of particles if the particle merging algorithm was not used. The equivalent number of photons and of pairs are the quantities of relevance. The cascade takes off around $\omega_0t = 30$ and the exponential trend appears to saturate around $\omega_0t = 45$. From this time forth, the number of pairs and photons still increases but rather following a secular growth. A very similar trend has also been observed by Nerush\cite{Nerush_laserlimit}. The characteristic time of absorption can be roughly estimated by equating the initial energy of laser pulses to the particle energy in the focal spot\cite{Nerush_laserlimit}, $\delta t \sim \ln(a_0n_c\sigma_0^2\tau_0c/n_0\lambda_0^3)/\Gamma = 19\omega_0^{-1}$  which is in good agreement with the value of the simulation: $\omega_0\delta t = 45-30 = 15$. The inset a) Fig. \ref{absortion_time} allows to establish the indisputable evidence that laser absorption becomes only visible at time $\omega t_a = 45$ when the relativistic critical density is reached (Fig. \ref{absorption_2d3d}) and that the depleted laser energy is almost all converted into energetic photons, whereas the energy contained in the pairs appears to be negligible. In order to fathom the sudden burst of gamma rays during the laser absorption, let us imagine a region of space filled with a standing wave and where the density of a plasma is rising with time. A comparable situation has been studied by Bulanov et al.\cite{Bulanov_damping} in the case of an electromagnetic wave propagating in a self-created pair plasma due to the Schwinger mechanism\cite{Schwinger}. The main conclusions of this previous work are that the wave is damped and the frequency of the wave is up-shifted due to in the increase of the plasma frequency. In our case, the standing wave created a pair plasma with a non uniform density. The density is maximum at the centre of the interaction zone and decreases symmetrically around this point, see Fig. \ref{absorption_2d3d}. When the density of the region around the centre of interaction becomes relativistically critical, the dynamics and the structure of the standing wave is broken. The two pulses, that are supposed to overlap in this region are partially damped in the critical zone and reflected outwards as pointed out in Fig. \ref{absorption_2d3d}-c. As the amplitude of the field goes down in the critical zone, the growth rate of the cascade is lessened as one notices in Fig. \ref{absortion_time}-b for $\omega_0t > 45$. The relativistic electrons and positrons forming the critical plasma wiggle in a disrupted wave and thus radiate hard photons that do not convert efficiently into pairs. The role of the gamma ray radiation can be seen as the equivalent of the role of the electron-ion collisions for the absorption of electromagnetic wave\cite{Mora}.

\subsection{Absorption efficiency}

The absorption efficiency can be directly seen on Fig. \ref{absortion_time}-c which shows the evolution of the laser energy as a function of time for different $a_0$. We now present a phenomenological model in order to understand the trend of the laser absorption. Let us consider two pulses with flat top envelop (the duration is $\tau_0$ and the amplitude $a_0$) counter propagating and initially in contact (but not overlapping) at $t=0$. The region where they collide is uniformly filled with seeded electrons of density $n_0$. In this one dimensional model, the absorption time is defined to be $n_0e^{\Gamma t_a} = a_0n_{c}$, i.e., $t_a = \ln(a_0n_c/n_0)/\Gamma$ with $\Gamma = \Gamma(a_0)$ During the collision of the two pulses, the density of the produced pair plasma reads
\begin{eqnarray}
\nonumber
n(x,t) &=& n_0\exp\left(\int_{|x|/c}^t \Gamma dt'\right) \\
&=& n_0\exp\left(\Gamma(t-|x|/c)\right),
\end{eqnarray}
for $|x| \leq ct$. The situation for a time of absorption bigger than half of the pulse duration is represented in Fig. \ref{cartoon}-a. The critical density is reached after the pulses have already fully overlapped and only the tail of the pulses can be efficiently absorbed. The characteristic width of the critical plasma created is $c(\tau_0-t_a)$. In the opposite situation, depicted in Fig. \ref{cartoon}-b, the pair plasma reaches the relativistic critical density soon after the collision, $t_a < \tau_0/2$. At $t=t_a$, a critical zone of characteristic width $ct_a$ is created. The two pulses cannot penetrate this zone and are reflected at the critical interface forming a new standing wave on both sides of the first critical zone. If the length of the pulse that is reflected is larger than $2ct_a$, a new critical pair plasma will be soon created, after a time $t_a$, and the process repeated until a small portion of the pulses is not wide enough to recreate a critical plasma. The typical pulse length leaving without interacting with the plasma is on the order of $ct_a$. Therefore, an initial length of $c(\tau_0-t_a)$ has been interacting with the critical plasma which is the same result as for $t_a > \tau_0/2$. We further assume that all parts of the pulse that have been interacting with the critical plasma will be eventually absorbed. The initial energy per unit surface contained in every pulse is $\mathcal{E}_L=E_0^2/4\pi c\tau_0$ and the energy absorbed is $\mathcal{E}_a=E_0^2/4\pi c(\tau_0-t_a)$. The laser absorption is then
\begin{equation}
\eta = \frac{\mathcal{E}_a}{\mathcal{E}_L}=1-\frac{t_a}{\tau_0},
\label{eq:absorption}
\end{equation}
with $t_a \leq \tau_0$. For the values of $a_0$, where $t_a > \tau_0$, the absorption is considered negligible and set to zero. Using the estimate for the growth rate of the cascade in case of linearly polarised lasers, Eq.(\ref{growthlinear}), we can evaluate the absorption efficiency as a function of $a_0$. The comparison between this model and the simulations results is plotted on Fig. \ref{cartoon}-c. In order to take into account the effect of higher dimensions and the characteristic parameters of the pulses, it is possible to rescale the time of absorption using the definition of the previous section, i.e., $t_a=\ln(a_0n_c\sigma_0^2\tau_0c/n_0\lambda_0^3)/\Gamma$. This latter estimate seems to be in fairly good agreement with the 2D results. In the simulation of Nerush\cite{Nerush_laserlimit}, the absorption time was found to be $t_a \simeq 10 \lambda_0/c$ and at time $t = 18 \lambda_0/c$ the authors observe $40\%$ of absorption, see Fig. 2 of Nerush\cite{Nerush_laserlimit}. By plugging these numbers in the simple formula Eq.(\ref{eq:absorption}), one obtains $\eta = 1-10/18 \simeq 0.44$ which is in good agreement with the numerical result. Due to the high computational cost of the 3D simulations to fully observe the absorption regime, we have only performed two simulations in full three-dimensional geometry for $a_0 = 1000$ and $a_0 = 2000$. Whereas the 3D results appear to be close from the 2D results of $a_0 = 2000$, the absorption appears to be overestimated in the 2D simulations for $a_0 = 1000$ which could be imputed to some side scattering effects in 3D.  
\begin{figure*}[t!]
\includegraphics[width=1\textwidth]{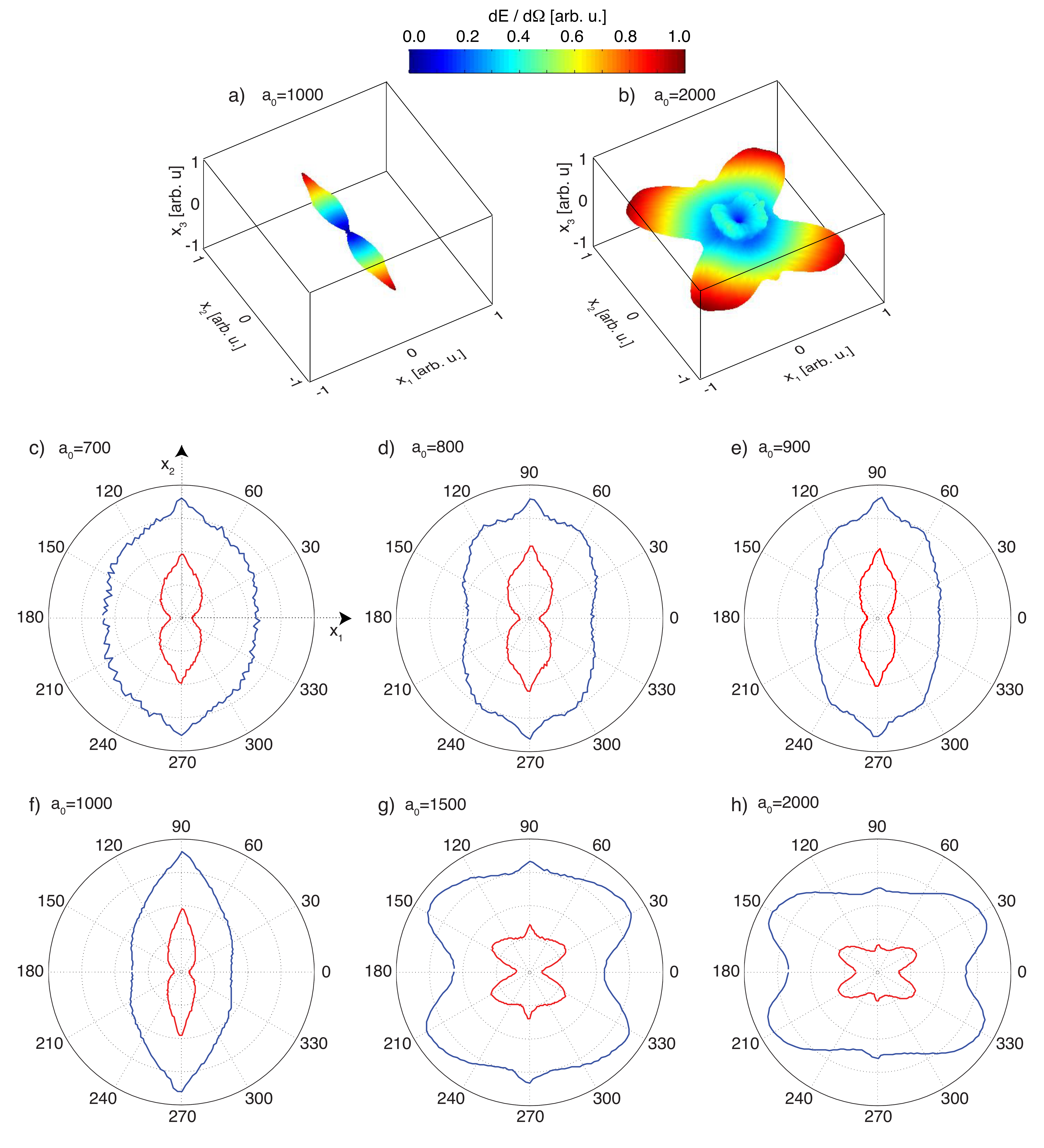}
\caption{Angular distribution of emitted radiation. 3D photon radiation maps from 3D simulations for a) $a_0=1000$ and b) $a_0=2000$ at $t=85~\omega_0^{-1}$. The radius from the centre of the box and colour are proportional to the amount of energy radiated per unit solid angle. c)-h) Polar radiation maps from 2D simulations all collected at $t=90~\omega_0^{-1}$. Radius is proportional to the amount of energy radiated per unit asimuthal angle.  Dark blue line corresponds to the photons above 2 MeV, while red is for the photons above 100 MeV.}
\label{radmap}
\end{figure*}

\subsection{Radiation map of Gamma-rays production}

It has been shown previously that quantum photon emission can generate $\gamma$-ray flashes \cite{Nakamura_2012, Gonoskov2014} and contribute for an efficient ultra-intense laser absorption \cite{Brady_2012, Ridgers_invited}. We now examine the gamma ray production in the scenarios we have explored to address laser absorption. Figure \ref{freq} shows radiation frequency spectra from 2D simulations at different laser intensities. The energy cutoff obtained is on on the GeV-level, which is consistent with the maximum energy the emitting electrons and positrons are able to develop in the cascade. Polar radiation maps for 2D simulations (Fig. \ref{radmap} c-h) show collected radiation above 2 MeV (dark blue) and above 100 MeV (red). It is readily evident for each example that the ÒwaistÓ of the polar map in the $x_1$-direction appears narrower for the high frequency radiation: this is because the high-energy photons emitted in the $x_1$ direction are more likely to reach a high $\chi_\gamma$ and decay into an electron positron pair than their counterparts emitted in different directions or at lower energies. The qualitative properties of radiation emission in our setup are dependent on the efficiency of laser absorption, i.e. the timing when the cascade reaches the relativistic critical density. While the pair plasma is relativistically underdense, the waves can propagate without losing a significant amount of energy. 

\begin{figure}
\includegraphics[width=0.5\textwidth]{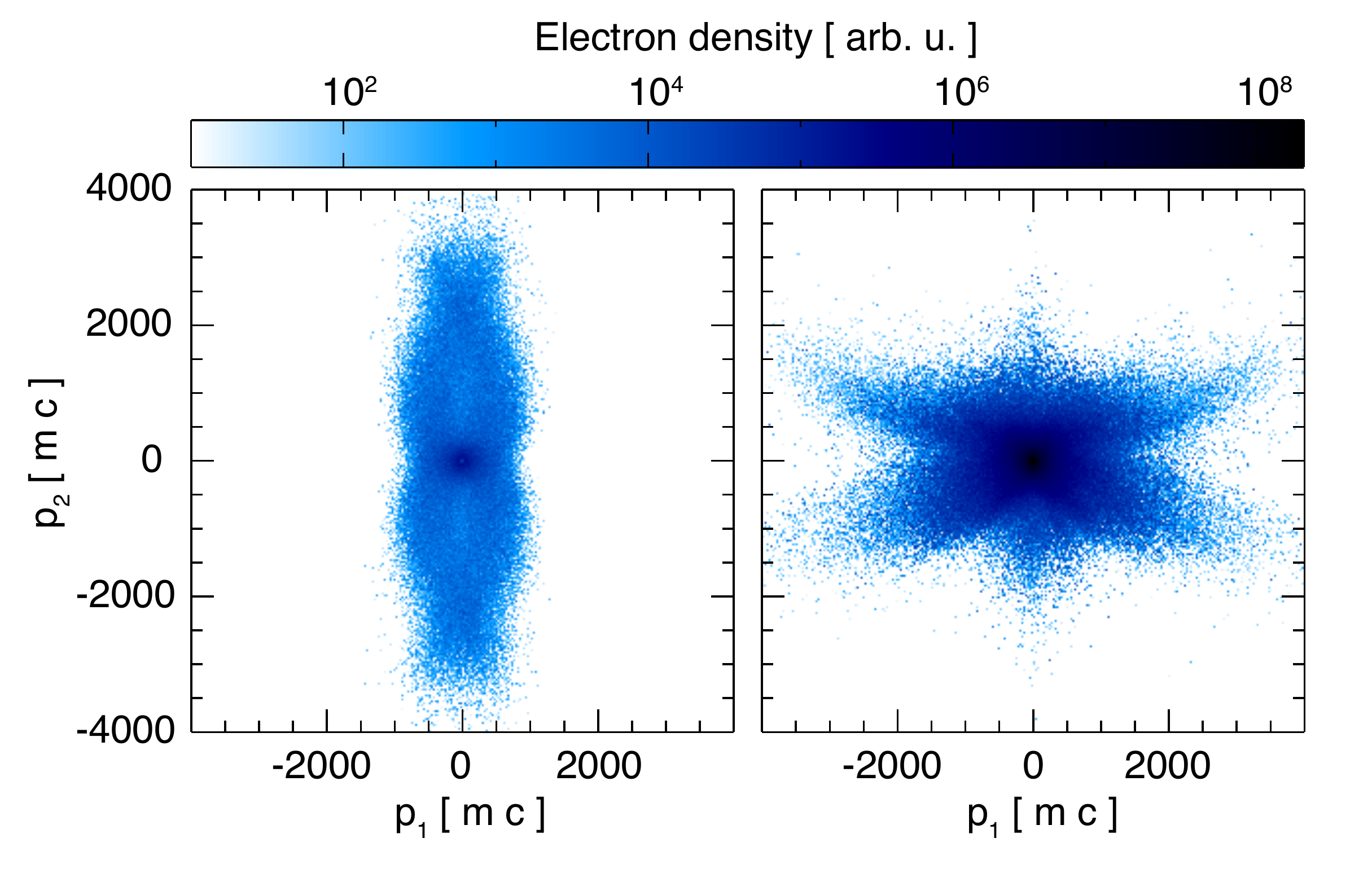}
\caption{Electron $p_1-p_2$ momentum phasespace at $t=65~\omega_0^{-1}$ for electrons within the interaction layer  $-2\lambda_0<x_1<2\lambda_0$. a) For $a_0=1000$ these electrons are within the sub-critical density plasma (critical density has not been reached yet in the simulation). b) For $a_0=2000$ these electrons form a relativistically critical layer }
\label{p2p1_ele}
\end{figure}

\begin{figure}
\includegraphics[width=0.5\textwidth]{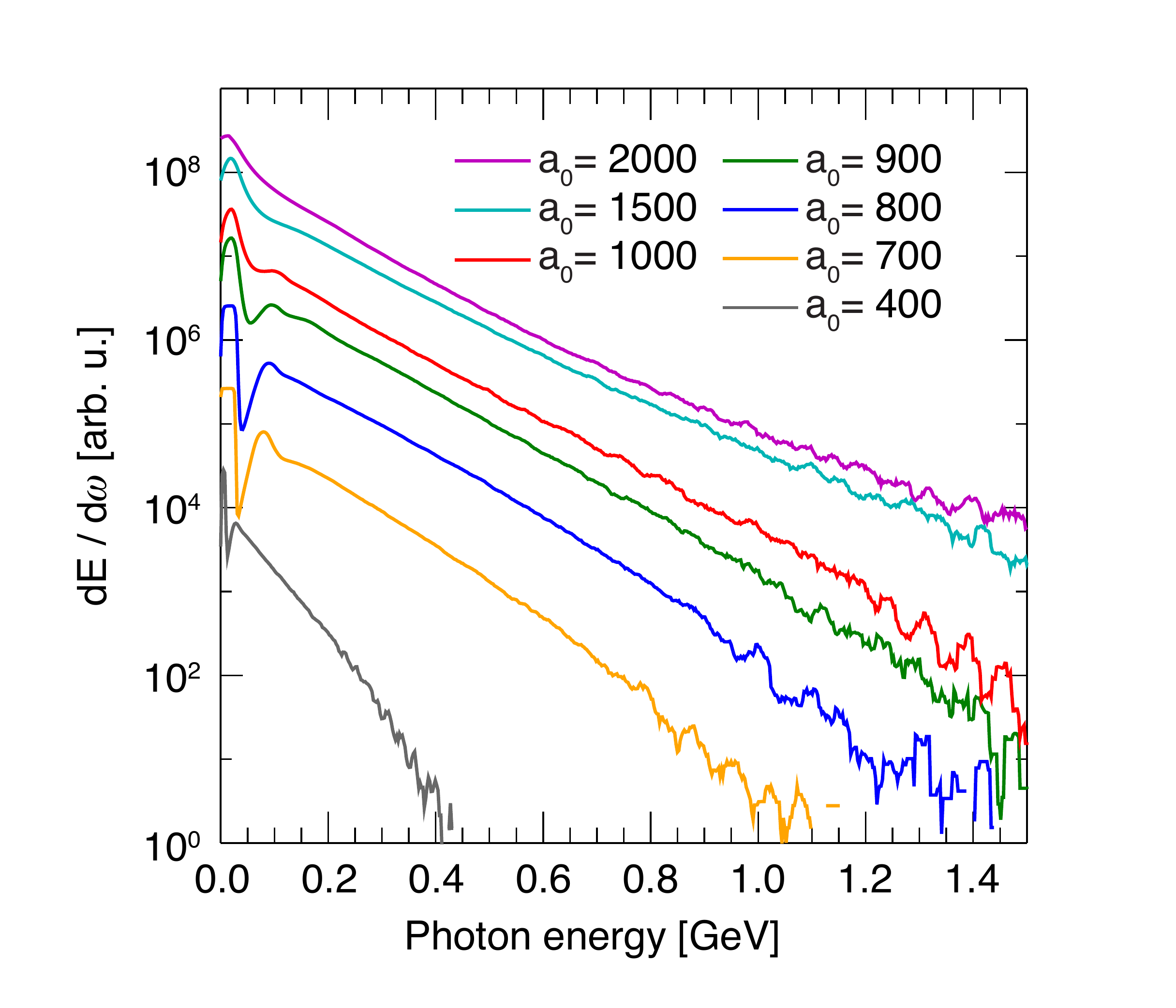}
\caption{Frequency spectra integrated in the solid angle for 2D simulations at different laser intensities.}
\label{gamma_spectra}
\end{figure}

For the case $t_a>\tau_0/2$ where we expect total laser absorption to be below 50 \%, the peak of the standing wave (created when the two pulses fully overlap) is not interacting with the critical plasma. This results in radiation emitted predominantly along the polarisation direction of the lasers $x_2$ (Fig. \ref{radmap} a, d-f), where the maximum particle momenta can be achieved (Fig.\ref{p2p1_ele}-a). In the over-critical region, the standing wave gets distorted, and eventually the trapped laser energy is efficiently converted into pairs and then $\gamma$-rays. If the relativistic critical density is reached before the lasers fully overlap ($t_a<\tau_0/2$), a large fraction of laser energy ($>$ 50 \%) is bound to be absorbed in the critical zone. Consequently, most energy conversion to photons will take place in the vicinity of this layer. More specifically, the layer is comprised of a core associated to the highest plasma density where the original standing wave is severely depleted and a surrounding area with near critical density where portions of progressive and standing waves (due to reflection) still exist. We show on Fig.\ref{p2p1_ele} the momentum phase space of the electrons around the absorption zone for $a_0=1000$ and $a_0=2000$. One notices the strong correlation between the typical pattern observed in the momentum phase space and in the radiation map which is due to the beaming effect of the radiation coming from ultra relativistic particles. The additional ÒcrossÓ pattern seen on Fig.\ref{p2p1_ele} for $a_0 = 2000$ is the signature of the copious amount of pairs quivering in the portions of progressive waves which also lead to the emission of energetic photons. The full radiation map is therefore the sum of the photons emitted during the absorption of the early standing wave and the subsequent absorption of the progressive waves begin reflected on different critical interfaces. If we compare the 2D and 3D simulations radiation maps for $a_0=1000$ and $a_0=2000$, the main features of the angular distribution of the radiated energy are perserved (c.f. Fig. \ref{radmap}b) indicating that this effect is not sensitive to the dimensionality of the problem. The angular distribution of the output radiation can therefore be used as a signature in experiments to verify if the critical density has been produced before or after the full laser overlap. As the latter is solely a function of laser parameters and the initial target seed, the output radiation could also serve as a possible diagnostic for the laser intensity achieved on-target. Finally, integrated spectra of emitted radiation from our 2D simulations at different laser intensities displayed in Fig.\ref{gamma_spectra} show the exponential feature at high energies and the energy cutoff on the GeV-level, which is consistent with the maximum energy the emitting electrons and positrons are able to develop while oscillating in the standing waves.

\section{Conclusion}

We have studied laser absorption in self-produced critical pair plasmas resulting from the saturation of QED cascades
for intensities relevant for near-future laser facilities such as ELI\cite{ELI} or the Vulcan 20 PW Project \cite{Vulcan}.
Seeded QED cascades are characterised by a single parameter, function of laser intensity and polarisation, that governs the exponential growth of the pairs and photons. As a rigorous analytical model for the growth rate in linearly polarised light is yet to be found, we have found, based on our previous study\cite{grismayer}, an empirical fit for the growth rate that is in good agreement with 2D and 3D QED-PIC simulations. An expression for the growth rate is proved to be a critical component to obtain a phenomenological model of laser absorption efficiency. The absorption model uses laser intensity, duration, and the seed plasma density to estimate the expected percentage of the depleted laser energy. This model has been thoroughly tested against our multi-dimensional simulation results with finite, diffraction-limited Gaussian laser pulses and against the results from the existing literature. Furthermore we also demonstrated the conditions to achieve substantial laser absorption, extending the pioneer results of previous works\cite{Nerush_laserlimit, Ridgers_invited, Brady_2012}. The consequence to be drawn is that the ratio between the absorption time (defined as the earliest time for the plasma to reach the relativistic critical density) and the pulse duration is the relevant parameter for the laser depletion.  If the relativistic critical density is achieved before the lasers fully overlap, the laser absorption should be higher than $50\%$. We have found that there is a qualitative difference in the hard photon polar radiation map when there is strong laser absorption ($>50\%$), compared with a case with low laser absorption ($<50\%$). This strong signature is not sensitive to the dimensionality of the problem (the qualitative features are the same in 2D and 3D), and can thus serve as a valuable diagnostic in future experiments. In addition to providing evidence for the existence of the relativistic-critical self-generated plasma, it can also be employed in experiments to evaluate the achieved laser intensity on target.  These qualitative signatures, along with the predictive capabilities of the absorption model, can serve as a valuable guide for the future experiments with next generation of lasers at ultra-high intensities. Further numerical and experimental work is now required to establish whether the gamma ray spectra occurring during these short-time cascades are compatible with the high energy photon spectra measured in the context of extreme astrophysical objects observations. 

\begin{acknowledgments}
The authors would like to thank Dr. Sebastian Meuren for stimulating and fecund discussions. This work is supported by the European Research Council (ERC-2010-AdG Grant 267841) and FCT (Portugal) Grants SFRH/BD/62137/2009 and SFRH/IF/01780/2013. We acknowledge PRACE for awarding access to resource SuperMUC based in Germany at Leibniz research centre. Simulations were performed at the Accelerates cluster (Lisbon, Portugal), and SuperMUC (Germany).
\end{acknowledgments}

%\bibliography{PoPinvited.bib}
%\bibliography{PoPinvited.bbl}
%Merlin.mbs v4.21 2009-07-09.
%

\end{document}